\documentclass[sn-mathphys-num]{sn-jnl}

\usepackage{graphicx}%
\usepackage{multirow}%
\usepackage{amsmath,amssymb,amsfonts}%
\usepackage{amsthm}%
\usepackage{mathrsfs}%
\usepackage[title]{appendix}%
\usepackage{xcolor}%
\usepackage{textcomp}%
\usepackage{manyfoot}%
\usepackage{booktabs}%
\usepackage{algorithm}%
\usepackage{algorithmicx}%
\usepackage{algpseudocode}%
\usepackage{listings}%

\let\oldAA\AA
\renewcommand{\AA}{\text{\normalfont\oldAA}}

\begin{document}

\title[Topological defects in a moir\'e pattern]{Higher order topological defects in a moir\'e lattice}


\author[1]{\fnm{Eugenio} \sur{Gambari}}
\author[2,3,4]{\fnm{Sebastian} \sur{Meyer}}
\author[1]{\fnm{Sacha} \sur{Guesne}}
\author[1]{\fnm{Pascal} \sur{David}}
\author[1]{\fnm{François} \sur{Debontridder}}
\author[5]{\fnm{Laurent} \sur{Limot}}
\author[5]{\fnm{Fabrice} \sur{Scheurer}}
\author[1]{\fnm{Christophe} \sur{Brun}}
\author[2,3,4]{\fnm{Bertrand} \sur{Dup\'e}}
\author[1]{\fnm{Tristan} \sur{Cren}}
\author*[1]{\fnm{Marie} \sur{Herv\'e}}\email{marie.herve@sorbonne-universite.fr}


\affil*[1]{\orgdiv{Sorbonne Universite, CNRS}, \orgname{Institut des Nanosciences de Paris, UMR7588}, \orgaddress{\street{4 place Jussieu}, \city{Paris}, \postcode{75005}, \country{France}}}
\affil[2]{\orgdiv{Universit\'e de Li\`ege}, \orgname{{Nanomat/Q-mat/CESAM}, \orgaddress{\street{Sart Tilman}, \city{City}, \postcode{B-4000}, \country{Belgium}}}}
\affil[3]{\orgdiv{Universit\'e de Li\`ege}, \orgname{{TOM/Q-mat/CESAM}, \orgaddress{\street{Sart Tilman}, \city{City}, \postcode{B-4000}, \country{Belgium}}}}
\affil[4]{\orgdiv{Fonds de la Recherche Scientifique (FRS-FNRS)}, \orgaddress{\city{Bruxelles},\country{Belgium}}}
\affil[5]{\orgdiv{Université de Strasbourg, CNRS}, \orgname{Institut de Phsyique et Chime des Matériaux de Strasbourg, UMR7504}, \orgaddress{\city{Strasbourg}, \postcode{F-67000}, \country{France}}}


\abstract{Topological defects are ubiquitous, they manifest in a wide variety of systems  such as liquid crystals, magnets or superconductors. The recent quest for non-abelian anyons in condensed matter physics stimulates the interest for topological defects since they can be hosted in vortices in quantum magnets or topological superconductors. In addition to these vortex defects, in this study we propose to investigate edge dislocations in 2D magnets as new building blocks for topological physics since they can be described as vortices in the structural phase field. Here we demonstrate the existence of higher order topological dislocations within the higher order moiré pattern of the van der Waals 2D magnet CrCl$_3$ deposited on Au(111). Surprizingly, these higher order dislocations arise from ordinary simple edge dislocations in the atomic lattice of CrCl$_3$. We provide a theoretical framework explaining the higher order dislocations as vortex with a winding Chern number of 2. We expect that these original defects could stabilize some anyons either in a 2D quantum magnet or within a 2D superconductor coupled to it. }



\maketitle

\section{Introduction}\label{sec1}

Van der Waals materials are emerging as extremely versatile building blocks for many fields of research, both at the fundamental and applied level \cite{Geim2013,Novoselov2016}. They offer tremendous possibilities for fine-tuning spintronics \cite{Wang2022,Han2014}, superconducting \cite{Kezilebieke2020,Kezilebieke2022,Martinez-Castro2023}, nanoelectronics \cite{Geim2013,Novoselov2016}, optical devices \cite{Basov2016,Zhang2021}. The toolbox of these heterostructures is continuously growing with for instance the recent discovery of ferromagnetic order in the family of chromium trihalides down to the monolayer limit, $\textrm{CrX}_\textrm{3}$ (X = I, Br, Cl) \cite{Huang2017,Chen2019,Kim2019,Soriano2020,Bedoya-pinto2021}. Due to the absence of magnetic anisotropy in the plane when the $\textrm{CrCl}_\textrm{3}$ monolayer is decoupled from its substrate it behaves as a 2D XY quantum magnet \cite{Bedoya-pinto2021} that can  host non-abelian anyons in the center of magnetic vortex cores \cite{kitaev2006}. Their integration in van der Waals heterostructure is expected to lead to a wealth of new exotic effects. Stacking chromium trihalides with transition metal dichalcogenides that exhibits a strong Ising spin-orbit coupling appears as a versatile and promising route towards the development of hybrid magnetic/superconducting systems \cite{Kezilebieke2020,Kezilebieke2022}. For instance, topological superconductivity has recently been shown to appear in $\textrm{CrBr}_\textrm{3}$/$\textrm{NbSe}_\textrm{2}$ heterostructures \cite{Kezilebieke2020,Kezilebieke2022}. The proposed mechanisms leading to a topological order in these hybrid structures was intimately related to the presence of a moir\'e pattern \cite{Kezilebieke2022}. Several studies have related the observation of a moir\'e structure to the mismatch between the $\textrm{CrX}_\textrm{3}$ monolayers and their substrate \cite{ganguli2023,Bedoya-pinto2021,Kezilebieke2022,Li2020}. The orientational degree of freedom may add a supplementary possibility to tune the moir\'e, as observed in twisted $\textrm{CrX}_\textrm{3}$ bilayers \cite{Xie2023,Qiu2021,Cheng2023}. The physics of moir\'e is quite complex, a same structure might exhibit several moir\'e patterns of different orders, as observed in Gr/Ir or Gr/Pt systems \cite{Zeller2014,Loginova2009,Zeller2012,Blanc2012,ndiaye2008,Merino2011,artaud2016}. Higher order structures appear when, in a particular direction, $n_1$ lattice periods of layer 1 almost match with $n_2$ periods of layer 2. Correspondingly, in the reciprocal space, a higher order moir\'e appears when a small enough wave vector $\textbf{q}_{moire}=m_1\textbf{q}_{1}-m_2\textbf{q}_{2}$ can be define with $\textbf{q}_i$  being the Bragg vectors of the two lattices and  $|\textbf{q}_{moire}|\ll \textrm{Min}(|\textbf{q}_1|,|\textbf{q}_2|)$. Ordinary moir\'e correspond to the case where $m_1=m_2=1$. Interestingly, the $\textrm{CrX}_\textrm{3}$ compounds have a unit cell almost twice the one of dense metals, that would naturally lead to a second order moir\'e where $m_1=1$ and $m_2=2$.

In the $\textrm{CrX}_\textrm{3}$ family, till now, the magnetisation was shown to be essentially collinear \cite{Tsubokawa1960,Starr1940,Huang2017,Chen2019,Bedoya-pinto2021}. However, in the $\textrm{CrI}_\textrm{3}$ bilayers, some hint of non-collinear magnetism was found. It has been related to the moir\'e pattern responsible of a non-negligible spatial modulation of the magnetic exchange interaction\cite{Xie2023,Cheng2023,Akram2021,Ghader2022,Fumega2023}. This confers an additional interest for the $\textrm{CrX}_\textrm{3}$ family. Indeed, a non-collinear magnetic order is predicted to be a major ingredient for inducing a topological order in superconducting/magnetic heterostructures \cite{Nakosai2013,Chen2015,Yang2016,Mohanta2019,Garnier2019}. Topological superconductors may host Majorana bound states localized on topological defects such as superconducting vortex cores or spin-orbit vortices \cite{Menard2019}, but other kind of defects such as dislocations could also be of interest since  edge dislocations are topological defects analogous to vortices \cite{Nye1974,dutreix2019}.  Interestingly, a moiré pattern in van der Waals materials seems to be a good candidate for exploring such physics since in graphene/Ir(111), graphene bilayers and h-BN/Ru \cite{coraux2008,Lu2014,Pochet2017,de_jong2022}, simple edge dislocations in the atomic lattice lead to dislocations in the moir\'e pattern.

Here we report that $\textrm{CrCl}_\textrm{3}$ deposited on a Au(111) substrate produces large two-dimensional islands that exhibit moir\'e pattern. In this particular case where the $\textrm{CrCl}_\textrm{3}$ lattice period is very close to twice the one of Au(111), the moir\'e pattern is of order 2. We show that the presence of an edge in the $\textrm{CrCl}_\textrm{3}$ lattice leads to a corresponding second order edge dislocation in the moir\'e. We demonstrate that this dislocation is a topological defect of Chern winding number of 2. These observations are explained within a unified model. We show that a higher order moir\'e pattern is a way to access to topological defects of higher Chern winding number. This unified model can be used as a predictive tool to build Van der Waals heterostructures in order to engineer topological defects of the desired Chern winding numbers.

\section{Results}\label{sec2}

\subsection{Superstructures in $\textrm{CrCl}_\textrm{3}$ monolayer on Au(111)}\label{subsec2}
\begin{figure}[h]
  \includegraphics[width=\linewidth]{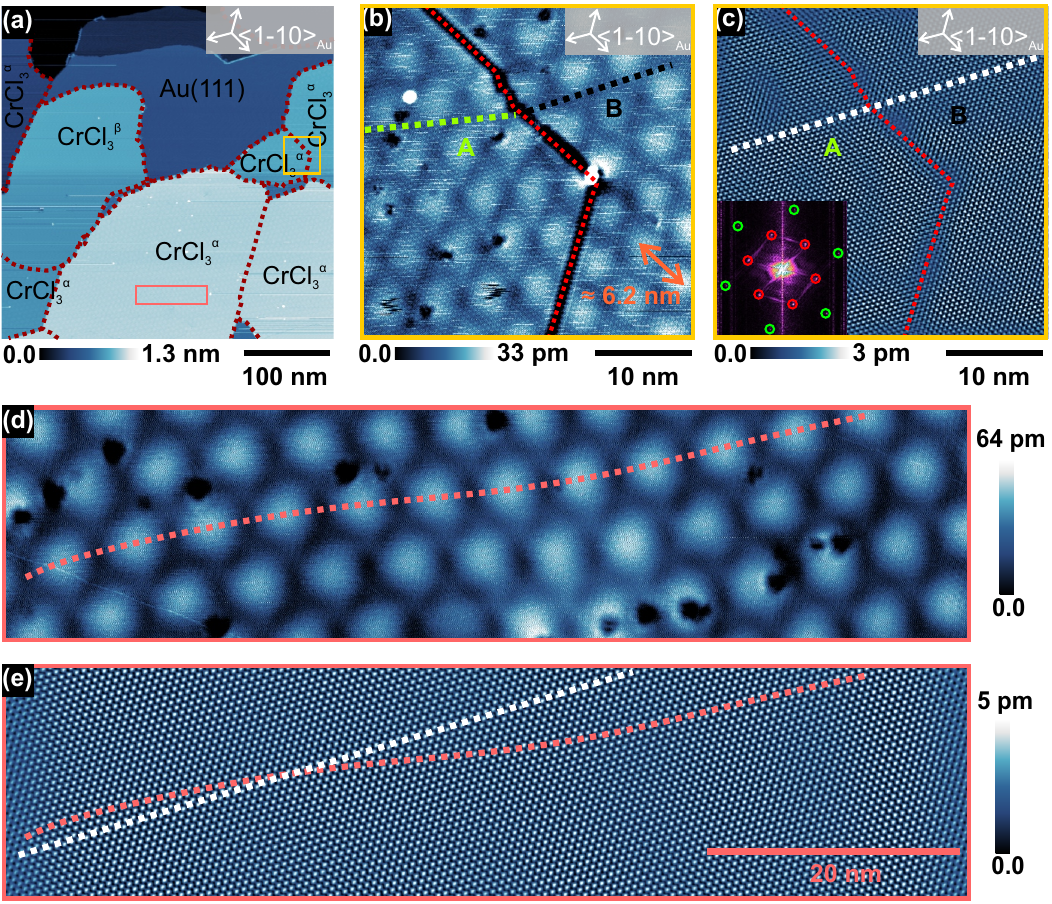}
  \caption{\textbf{Superstructure in $\textrm{CrCl}_\textrm{3}$/Au(111).} (a) Large scale STM topography of 0.6 monolayer of $\textrm{CrCl}_\textrm{3}$ deposited on a Au(111) substrate. (b) Zoom in the yellow frame of (a) showing 2 structural domains where the superstructure orientation differs by an angle of 13°. The green and dark dashed lines are guides for the eyes to follow the orientation of the superstructure in the two domains, the red dashed line shows the separation between the two islands. (c) Fourier filtered image revealing the atomic lattice of  $\textrm{CrCl}_\textrm{3}$ in the two structural domains. The white dashed line is a guide for the eye to follow the orientation of the atomic lattice. (d) STM topography acquired in the pink frame of (a) showing that the superstructure is undulating inside a $\textrm{CrCl}_\textrm{3}$ island, while no rotation of the atomic lattice can be detected in the Fourier filtered image (e). The pink and white dashed lines are eye-guides to follow the orientation of the superstructure and the atomic lattice inside the island. For all the STM images, the tunneling parameters are: Bias voltage, $\textrm{U}_\textrm{0}=2$~V; Tunneling current,  $\textrm{I}_\textrm{t}=200$~pA}
  \label{fig1}
\end{figure}

Crystalline chromium trichloride ($\textrm{CrCl}_\textrm{3}$) thin films were grown using molecular beam epitaxy (MBE) on a clean Au(111) substrate. Figure~\ref{fig1}.a depicts a large scale STM topography of 0.6 monolayers (ML) of $\textrm{CrCl}_\textrm{3}$ measured at 4~K. The dashed red line highlights the presence of 1~ML thick islands of $\textrm{CrCl}_\textrm{3}$. Two distinct types of islands denoted as $\textrm{CrCl}_\textrm{3}^\alpha$ and $\textrm{CrCl}_\textrm{3}^\beta$ were observed. Notably, these islands display two different preferred in-plane crystalline orientations rotated by 30° with respect to each other (see Supplementary note and figure S1.1 and S1.2) .

Throughout the manuscript, we will exclusively focus on discussing the $\textrm{CrCl}_\textrm{3}^\alpha$ islands. In these islands, the dense atomic directions of $\textrm{CrCl}_\textrm{3}$ aligns along the dense atomic directions of the Au substrate, the $\textrm{$<$1$\bar{1}$0$>$}_\textrm{Au}$ direction. In these islands, our STM study shows that  $\textrm{CrCl}_\textrm{3}$ is restoring the lattice parameter of a free standing  $\textrm{CrCl}_\textrm{3}$ monolayer (see Supplementary note S1.1 and Supplementary figure S1.1 and S1.2 ). Remarkably, upon cooling down the sample to 4K, a superstructure emerges within these $\textrm{CrCl}_\textrm{3}^\alpha$ islands, which is not visible at room temperature (see Supplementary figure S2). Figure~\ref{fig1}.b shows an STM image featuring two islands labelled A and B, which form a twin boundary (see Supplementary figure S4). Both islands exhibit an hexagonal superstructure with a periodicity of 6.2 nm. Their orientation differ by an angle of 13° (see green and dark dashed lines). Interestingly, the rotation of the superstructure does not seem, at first sight, to be correlated with a rotation of the atomic lattice. Figure~\ref{fig1}.c presents the Fourier filtering of the STM topography of figure ~\ref{fig1}.b revealing the atomic structure of $\textrm{CrCl}_\textrm{3}$. Fourier filtering is obtained by selecting the Bragg spots marked in red and green in the fast Fourier transform (FFT) of Figure~\ref{fig1}.c. The white dashed line serves as a visual guide, highlighting the dense atomic direction of $\textrm{CrCl}_\textrm{3}$ in both islands. At first sight, no misalignment of the crystal axis is detectable between these two islands. Additionally, within individual islands, the superstructure can exhibit a distortion, while the atomic lattice seems to remain unaffected. This is shown in Figure~\ref{fig1}.d and ~\ref{fig1}.e. Figure~\ref{fig1}.d displays an STM topography acquired inside an island. The pink dashed line is a guide for the eye showing that the superstructure is not straight but slightly undulate. Figure~\ref{fig1}.e shows the corresponding Fourier filtered image, revealing the atomic lattice of $\textrm{CrCl}_\textrm{3}$. The white dashed line that follows the dense atomic direction seems to be insensitive to the change of orientation in the superstructure. Nevertheless, we conducted modelization of a moiré pattern in this system (see Supplementary note and figure S5) and show that this superstructure is indeed a moiré effect which shows an unusual high sensitivity to subtle variations in the atomic lattice orientation. The change of orientation of the moiré pattern shown in Figure~\ref{fig1}.b and d is explained by a deviation angle in the atomic lattice of less than 1°. This system is rather rich with plenty of different moiré patterns induced by very little deviations of the  $\textrm{CrCl}_\textrm{3}$ lattice with respect to Au(111). It could be a good plateform to explore the influence of a moiré on the magnetic texture.  

\subsection{Role of Au(111) on the electronic structure}\label{subsec2}

\begin{figure}[h]
  \includegraphics[width=\linewidth]{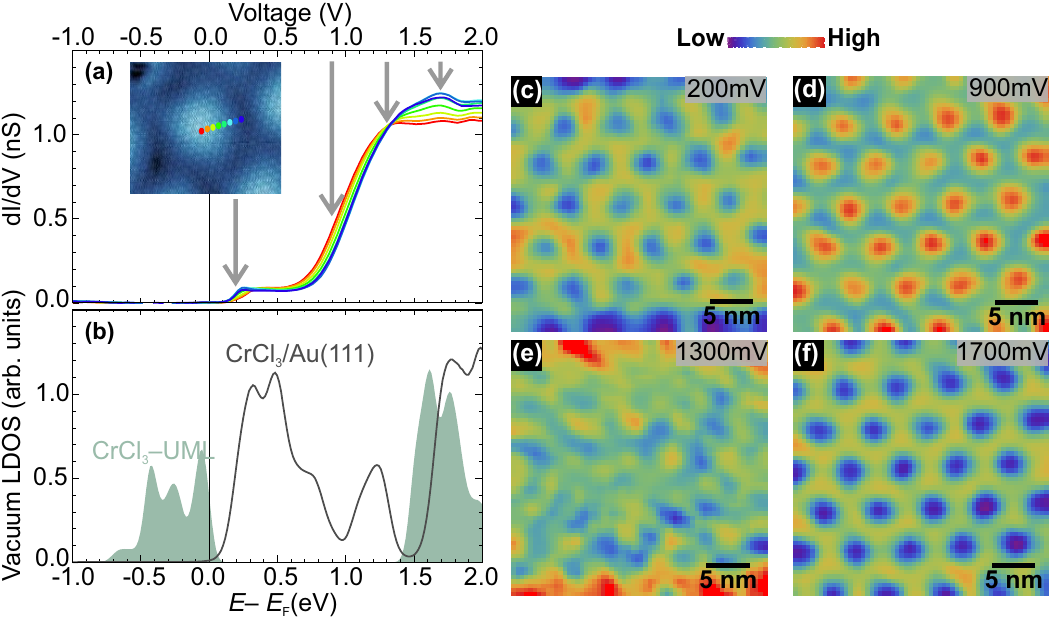}
  \caption{\textbf{Role of Au(111) on the electronic structure.} Comparison between experimentally measured spectroscopy data and calculated vacuum density of states of CrCl\textsubscript{3}/Au(111). (a) Scanning tunneling spectroscopy measurements on different positions of the moir\'e pattern as indicated in the inset. (c), (d), (e) and (f) dI/dV mapping of the moir\'e pattern taken at 4 different energies. (setpoint: $\textrm{U}_\textrm{0}=2$~V,  $\textrm{I}_\textrm{t}=400$~pA, these data were recorded using a lock in, a modulation of 20mV rms was added to the bias voltage, the full spectroscopic map was recorded from 2 V to -1 V) ((b) Density functional theory calculations of the vacuum density of states 3 \AA~above the surface of a CrCl\textsubscript{3} unsupported monolayer (UML, green area) in comparison with CrCl\textsubscript{3}/Au(111) (black line). Note that in the calculations, it is assumed that CrCl\textsubscript{3} is in perfect registry with Au(111), that is the moiré is not taken into account.}
  \label{fig: spectroscopy}
\end{figure}

From our measurements, we deduce that CrCl\textsubscript{3} restores its lattice constant as in monolayer without substrate (see Supplementary note S1.1 and Supplementary figure S1.2), which is why we need to clarify the role of Au(111) on the electronic structure of the system. Scanning tunneling microscopy imaging reflects the local density of states and does not reflect directly the structure of the surfaces. It contains also some information on the local electronic structure that can eventually be modulated along the moir\'e pattern.
We have performed tunneling spectroscopy experiments in order to elucidate the origin of the observed corrugation in the topographic measurements. Figure~\ref{fig: spectroscopy}.a shows scanning tunneling spectroscopy data for different positions above the moir\'e pattern of CrCl\textsubscript{3}/Au(111) (from red, center, to blue, edge of the pattern). Regardless of the position, the measured conductance shows the same features, that is, no intensity below 0.2 eV; then there is a small increase at 0.2 eV followed by a plateau and then a strong increase after 0.3 eV, eventually followed by a plateau above ~1.2 eV. The behavior of the conductance above 1.2 eV is strongly dependent on the tip position over the moiré pattern.  Figure~\ref{fig: spectroscopy}.c to f display 4 conductance maps revealing the spatial variation of the local density of states over the moir\'e. Figure~\ref{fig: spectroscopy}.c, recorded at 0.2 eV, shows a pattern with a lower density of states in the center of the protrusion found in the topography sector (Figure~\ref{fig: spectroscopy}.a). At 0.9 eV, the contrast is reversed, the highest density of state being in center of the protrusions (Figure~\ref{fig: spectroscopy}.d). At $1.3$~eV the contrast cancels (Figure~\ref{fig: spectroscopy}.e), it can be seen in the tunneling spectra of Figure~\ref{fig: spectroscopy}.a that it corresponds to a crossing point where all spectra cross. Above  $1.3$~eV a second contrast inversion is observed with a lower density of states in center over the protrusions (Figure~\ref{fig: spectroscopy}.f). These measurements show that the moir\'e pattern produces quite clear spectroscopic features and thus we expect a strong effect due to the structural modulation of the moir\'e on the electronic and magnetic properties of the CrCl$_3$ layer. This is corroborated by the fact that the moir\'e pattern is not showing up in the topographic images at room temperature. We have performed a temperature ramp from 4~K to 130~K and we observed a gradual decrease of the moir\'e corrugation as the temperature is increased. We attribute this to the fact that the moir\'e seen in the topography is mainly of spectroscopic origin, i.e. there is  no real corrugation (see Supplementary note and figure S3). The fact that the moir\'e reflects mainly a modulation of the local density of states suggests the importance to investigate the electronic band structure by density functional theory.

We therefore calculate the vacuum density of states (VacDOS) following the Tersoff \& Hamann model \cite{Bardeen1961,Tersoff-Hamann1985,Wortmann2001} to compare and interpret the spectroscopy data. In Figure~\ref{fig: spectroscopy}.b, we see the results for an unsupported CrCl\textsubscript{3} monolayer (UML) as the green shaded area (for further information, see the Methods section and Supplementary note and figure S6). Without the Au(111) substrate, we would expect peaks to occur at the Fermi energy and more significantly below the Fermi energy, where the experimental data do not show any states. A gap appears in between $0.2$~eV and $1.4$~eV, which is absent in the measurement of Fig.~\ref{fig: spectroscopy} (a). When adding Au underneath of CrCl\textsubscript{3}, the VacDOS changes drastically, shifting the band gap to the Fermi level and below (black curve). Note that due to the computational effort within the DFT calculation, we adapt a perfect registry between CrCl\textsubscript{3} and Au(111). Nevertheless, the calculated VacDOS (Figure~\ref{fig: spectroscopy}.b - dark curve) exhibits qualitatively the same trends as in the measurement (cf. panel (a) and (b)): the band gap and the shoulder/peak structures at 0.2~eV and 1.7~eV. We attribute the differences between the calculations and the measurements to the differing structures within the calculations.
From this comparison, we deduce that Au has an influence on the electronic structure of CrCl\textsubscript{3} and that this system is not solely driven by Van-der Waals forces. This is in accordance with reference \cite{Zhang2022}, where the authors compared the binding energies and concluded that for CrCl\textsubscript{3}/Au(111) these energies lie at the threshold of VdW and ionic bonds.

\subsection{Higher order edge dislocations of the moir\'e pattern}\label{subsec2}
\begin{figure}[h]
  \includegraphics[width=\linewidth]{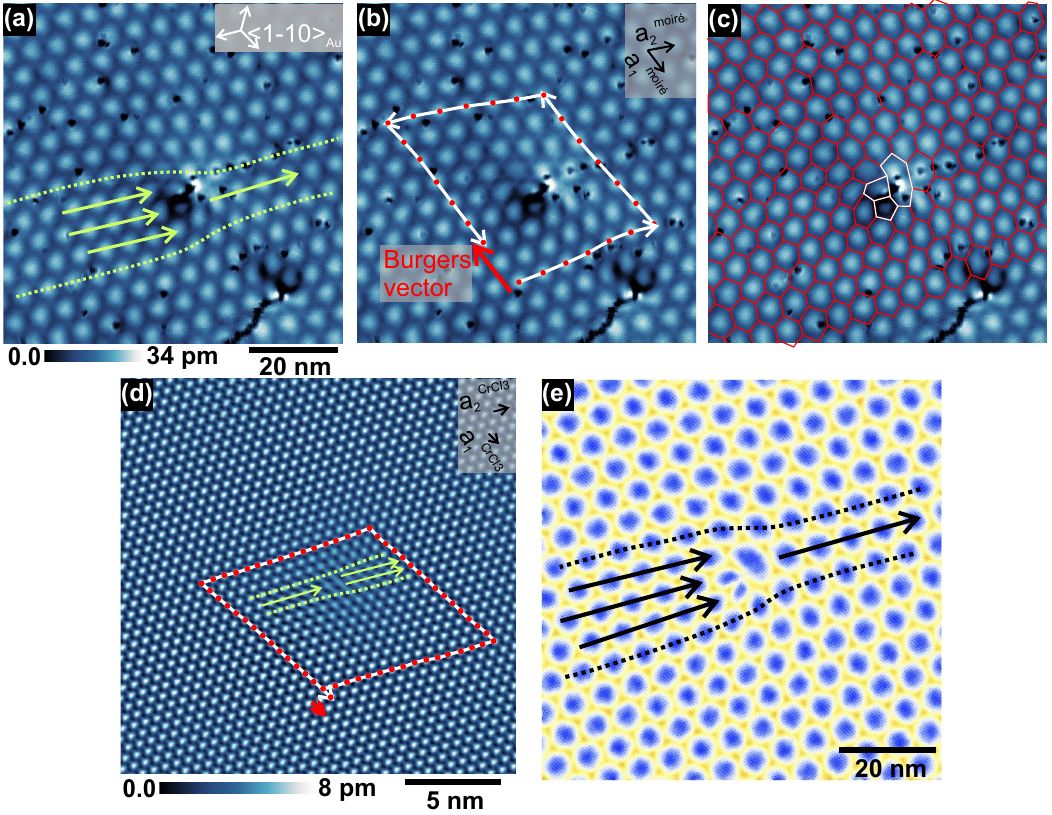}
  \caption{\textbf{Higher order edge dislocations.} (a) STM topograhy showing a double dislocation in the moir\'e pattern. The dashed green lines and arrows are guide for the eyes to follow the dislocation. ($\textrm{U}_\textrm{0}=2$~V,  $\textrm{I}_\textrm{t}=200$~pA) (b) Drawing of the Burgers circuit around the dislocation, and its coordination (c). (d) Zoom around the dislocation in the moir\'e pattern. The STM topography was Fourier filtered to reveal the atomic lattice. It shows a simple dislocation in the atomic lattice of  $\textrm{CrCl}_\textrm{3}$ at the exact position of the dislocation in the moir\'e pattern. (e) Numerical simulation of the dislocation in the moir\'e pattern.}
  \label{fig3}
\end{figure}

Very peculiar defects have been observed in this moiré pattern. They consist in higher order edge dislocations. One of those particular defect is shown in Figure~\ref{fig3}.a. On the left side of the dislocation, as indicated by the green arrows, three rows of the moir\'e pattern merge into a single row on the right side, resulting in the disappearance of a pair of rows. Notably, we have observed dislocations with similar characteristics in multiple locations of the sample. For each dislocation, a pair of rows vanishes, which represents an inherent characteristic of this particular moir\'e pattern.

To further characterize the dislocation, we analyze its Burgers vector (Figure~\ref{fig3}.b) and coordination (Figure~\ref{fig3}.c). The Burgers vector of the dislocation is found to be twice the lattice vector of the moir\'e pattern: $\boldsymbol{b}_{moir\acute{e}} = -2\boldsymbol{a}_{moir\acute{e}}^{1}$. The coordination of the lattice cells surrounding the dislocation is determined to be 5-5-8, i.e. two cells are five fold coordinated while one is height fold, this is in contrast with ordinary 5-7 dislocations. To gain insight into the atomic structure surrounding the dislocation, we provide a magnified view around the dislocation in the STM topography (Figure~\ref{fig3}.d). By Fourier filtering the Bragg spots, the atomic lattice of $\textrm{CrCl}_\textrm{3}$ surrounding the dislocation becomes visible. Remarkably, at the exact position of the dislocation in the moir\'e pattern, we also observe an edge dislocation in the atomic lattice of $\textrm{CrCl}_\textrm{3}$. Specifically, one row of atoms on the left side of the dislocation transforms into two rows on the right side, as indicated by the green arrows. Prior studies in the literature has suggested that a dislocation or defect within a moir\'e pattern faithfully replicates the corresponding dislocation or defect at the atomic level \cite{Lu2014,Pochet2017,de_jong2022,yang2022}. However, in this case, the dislocation at the atomic level is characterized by a Burgers vector $\boldsymbol{b}_{{CrCl}_{3}} = \boldsymbol{a}_{{CrCl}_{3}}^{1}$ (as shown by the red arrow in Figure~\ref{fig3}.d) and a 5-7 defect. This is in contrast, with the dislocation in the moir\'e pattern where the Burgers vector is opposite to the one in the atomic lattice and 2 times larger than the moiré lattice vector.  

To understand the particularity of this dislocation in the moir\'e pattern, we modelize our system by introducing a 5-7 defect in the atomic lattice of $\textrm{CrCl}_\textrm{3}$ and convolute it with a defect-free atomic lattice of Au(111). Figure~\ref{fig3}.e illustrates the result of this simulation, which successfully reproduces the experimental observations of a dislocation with a Burgers vector $\boldsymbol{b}_{moir\acute{e}} = -2\boldsymbol{a}_{moir\acute{e}}^{1}$.

\subsection{Analytical description}\label{subsec2}
To understand the physical origin of the observed moir\'e and its peculiar dislocations, we conducted an analytical description of the moir\'e pattern. A conventional approach to describe a moir\'e pattern involves employing a continuous model \cite{Zeller2014}. The moir\'e may be described as the product between two lattice functions, denoted as $\textrm{f}_\textrm{1}(x,y)$ and $\textrm{f}_\textrm{2}(x,y)$ which are the superposition of plane waves on top of a constant background with an associated spatial frequency $Q_{1}$  and $Q_{2}$ respectively. To illustrate this with the simplest case, we consider the scenario of waves in each lattice propagating along the x-direction. The product between two lattice functions $f_{1}(x,y)\times f_{2}(x,y) =\left(1+a\cos(Q_{1}x))\times(1+a\cos(Q_{2}x)\right)$ can be rewritten as:

\begin{equation}
f_{1}\times f_{2} =1+\frac{a^2}{2}\left(\cos[(Q_{1}+Q_{2})x]+\cos[(Q_{1}-Q_{2})x]\right)+2a \cos[\frac{(Q_{1}+Q_{2})}{2}x] \cos[\frac{(Q_{1}-Q_{2})}{2}x]
\label{eq1}
\end{equation}

We obtain the addition of 3 terms: a constant background, and 2 periodic function. In the following, to describe the moiré we will consider the first periodic function. It is a sum of two cosines with one slowly varying term $\cos[(Q_{1}-Q_{2})x]$ that stems for the moir\'e term. The resulting spatial beating frequency, $Q_{1}-Q_{2}$ corresponds to the spatial frequency of the moir\'e, $Q_{moire}$.  For the case of a moir\'e formed between two hexagonal lattices as for $\textrm{CrCl}_\textrm{3}$/Au(111), each lattice needs to be treated as the superposition of two waves propagating in directions at 120° one from the other. This renders the mathematical description in the real space heavy. A simpler way to describe a moir\'e is to treat the problem in the reciprocal space where the moir\'e wave vector is expressed as the difference between the wave vectors of each atomic lattice. We employed the intelligible model described by Zeller et al. and Le Ster et al. \cite{Zeller2014,ster2019}. Within this framework, not only the 1st order, but also the higher order Fourier component of the moir\'e need to be considered.

The general reciprocal vectors for both the Au and $\textrm{CrCl}_\textrm{3}$ atomic lattices, $\boldsymbol{Q}_{Au}^{i,j}$  and $\boldsymbol{Q}_{{CrCl}_{3}}^{k,l}$ are expressed as linear combinations of the primitive reciprocal lattice vectors, namely $\boldsymbol{Q}_{Au}^{1,0}$, $\boldsymbol{Q}_{Au}^{0,1}$, $\boldsymbol{Q}_{{CrCl}_{3}}^{1,0}$, $\boldsymbol{Q}_{{CrCl}_{3}}^{0,1}$: 

\begin{align}
\boldsymbol{Q}_{Au}^{i,j} &=i\boldsymbol{Q}_{Au}^{1,0}+j\boldsymbol{Q}_{Au}^{0,1}\\
\boldsymbol{Q}_{{CrCl}_{3}}^{k,l} &=k\boldsymbol{Q}_{{CrCl}_{3}}^{1,0}+l\boldsymbol{Q}_{{CrCl}_{3}}^{0,1}.
\end{align}
\begin{figure}[h]
  \includegraphics[width=11.6cm]{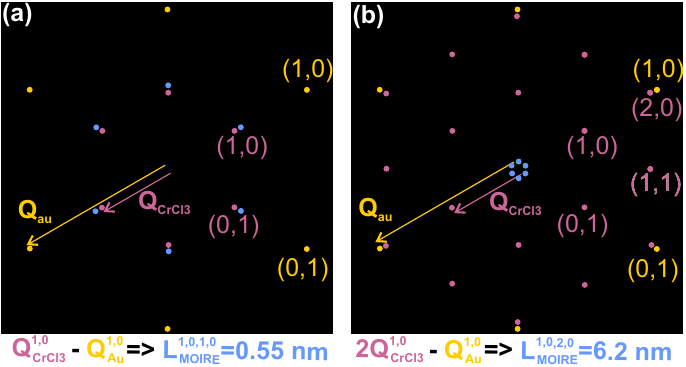}
  \caption{\textbf{Reciprocal lattice of $\textrm{CrCl}_\textrm{3}$/Au(111).} The blue dots corresponds to the 1st (a) and 2nd (b) order moir\'e spots. }
  \label{fig4}
\end{figure}

Here, $|\boldsymbol{Q}_{Au}^{1,0}|$ $=|\boldsymbol{Q}_{Au}^{0,1}|=2\pi/a_{Au}$ and $|\boldsymbol{Q}_{{CrCl}_{3}}^{1,0}|=|\boldsymbol{Q}_{{CrCl}_{3}}^{0,1}|=2\pi/a_{{CrCl}_{3}}$, where ${a_{Au}}= 0.288$~nm and $a_{{CrCl}_{3}}=0.604$~nm corresponds to the lattice parameters of Au and $\textrm{CrCl}_\textrm{3}$, respectively. The possible Fourier components of the moir\'e pattern's reciprocal vectors are expressed as the difference between the corresponding reciprocal vectors of Au and $\textrm{CrCl}_\textrm{3}$, leading to:
\begin{equation}
\boldsymbol{Q}_{moir\acute{e}}^{i,j,k,l}=\boldsymbol{Q}_{Au}^{i,j}-\boldsymbol{Q}_{{CrCl}_{3}}^{k,l}.
\end{equation}
When the dense atomic directions of $\textrm{CrCl}_\textrm{3}$ align with the $\textrm{$<$1$\bar{1}$0$>$}_\textrm{Au}$ directions, $\boldsymbol{Q}_{Au}^{1,0}$ is colinear with $\boldsymbol{Q}_{{CrCl}_{3}}^{1,0}$, as depicted in Figure~\ref{fig4}.a. Let us illustrate why one should here consider higher order moir\'e by first showing that a usual first order moir\'e is doesn't correspond to our observation. The first order moir\'e wave vector formed between the (1,0) Bragg spots of Au and $\textrm{CrCl}_\textrm{3}$ is expressed as:

\begin{equation}
\boldsymbol{Q}_{moir\acute{e}}^{1,0,1,0}=\boldsymbol{Q}_{Au}^{1,0}-\boldsymbol{Q}_{{CrCl}_{3}}^{1,0}
\end{equation}

This results in a moir\'e period in real space: 

\begin{equation}
L_{moir\acute{e}}^{1,0,1,0}=2\pi/|\boldsymbol{Q}_{moir\acute{e}^{1,0,1,0}}|=0.55~\textrm{nm}.
\end{equation}

This period is notably smaller than 0.604 nm, the lattice parameter of  $\textrm{CrCl}_\textrm{3}$. As a result, a moir\'e pattern between the (1,0) Bragg spots of Au and $\textrm{CrCl}_\textrm{3}$ is meaningless. This emanates from the fact that $a_{{CrCl}_{3}}$ exceeds the lattice parameter of Au by more than a factor two. Noteworthy, $\textrm{CrCl}_\textrm{3}$/Au(111) is the paragon case where higher order moir\'e is popping up. Extending the reciprocal lattice of $\textrm{CrCl}_\textrm{3}$ to the second order (as depicted in Figure~\ref{fig4}.b) reveals that the shortest moir\'e wave vector is formed between the (1,0) Bragg spot of Au and the (2,0) Bragg spot of $\textrm{CrCl}_\textrm{3}$, expressed as: 

\begin{equation}
\boldsymbol{Q}_{moir\acute{e}}^{1,0,2,0}=\boldsymbol{Q}_{Au}^{1,0}-\boldsymbol{Q}_{{CrCl}_{3}}^{2,0}.
\end{equation}

In the real space, this yields a moir\'e pattern period:

\begin{equation}
L_{moir\acute{e}}^{1,0,2,0}=2\pi/|\boldsymbol{Q}_{moir\acute{e}^{1,0,2,0}}|=6.2~\text{nm},
\end{equation}

which corresponds exactly to the period we measure in our experiments.

\begin{figure}[h]
  \includegraphics[width=\linewidth]{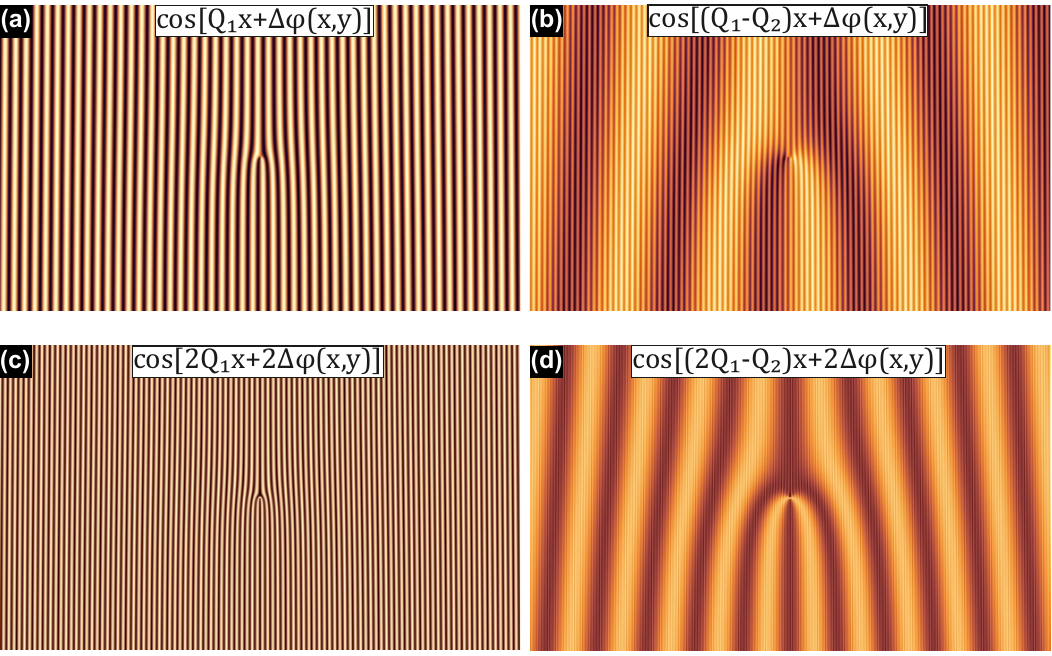}
  \caption{\textbf{Analytical description of the dislocation in the moir\'e pattern.} (a) First order periodic lattice simulated by a simple plane wave in which a dislocation has been introduced in the form of phase difference which correspond to a Berry phase: $\Delta\varphi(x,y)$. (b) First order moir\'e pattern when a simple dislocation exists in one of the two periodic lattices: a simple dislocation appears in the moir\'e. The dislocation in the moiré pattern exhibits the same Berry phase as in the primary lattice. (c) Second order primary lattice: the Berry phase of the dislocation is doubled ($2\Delta\varphi(x,y)$) with respect to the first order shown in (a).  (d) The resulting second order moiré pattern shows an additional double row with a Berry phase ($2\Delta\varphi(x,y)$) .} 
  \label{fig5}
\end{figure}

We will now explain the emergence of the peculiar dislocations (Figure~\ref{fig3}).  As a dislocation is a punctual defect that cannot be modeled easily in reciprocal space, the problem needs to be partially treated in real space. In order to avoid a heavy mathematical description and maintain conceptual clarity, we will consider only a dislocation in a simple planar wave that can be modeled by the expression $\cos[(Q_{1}x+\Delta\varphi(x,y)]$, with $\Delta\varphi(x,y)$ being a Berry phase associated to the defect. If one considers a dislocation located at the origin, $\Delta\varphi(x,y)$ can just be taken as the angle of the position vector $\boldsymbol{r}=(x,y)$ to the abscissa, i.e.  $\Delta\varphi(x,y)=\arg(x+i y)$. The resulting dislocation pattern is shown in Figure~\ref{fig5}.a.

Consider now two lattices forming a first order moir\'e, with a dislocation in one of the lattices. It naturally results in a dislocation in the moir\'e pattern, as shown in Figure~\ref{fig5}.b. This can easily be explained using eq.\ref{eq1}. The moir\'e term is $\cos[((Q_1-Q_2)x+\Delta\varphi(x,y)]$ which also exhibits a Berry phase that naturally gives a dislocation in the moir\'e pattern. The question that immediately arises is what about our higher order moir\'e? The second order moir\'e is a beating between $\cos(2 Q_1x)$ and $\cos(Q_2x)$, however, in the presence of a dislocation the phase $Q_1x$ has to be replaced by $(Q_{1}x+\Delta\varphi(x,y))$. This results in a beating between $\cos[2(Q_{1}x+\Delta\varphi(x,y))]$ and $\cos(Q_2x)$ that leads to a moir\'e term $\cos[(2Q_1-Q_2)x+2\Delta\varphi(x,y)]$. As we can see, the resulting Berry phase is now $2\Delta\varphi(x,y)$, the double of the Berry phase in the primary defect. Figure~\ref{fig5}.c,d shows the resulting dislocation patterns in the second order lattice and its moir\'e pattern.

\begin{figure}[h]
  \includegraphics[width=\linewidth]{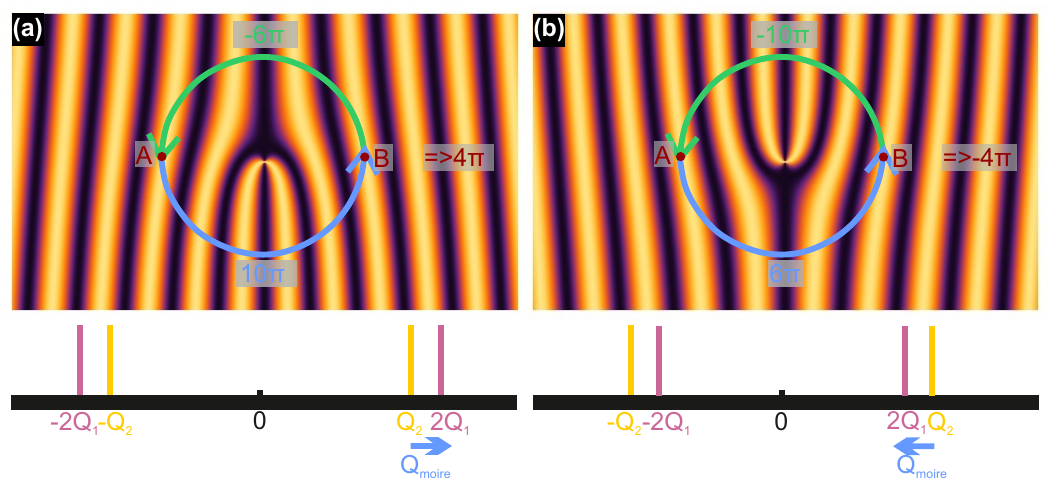}
  \caption{\textbf{Dislocation's Berry phase.} Dislocation in the 2nd order moir\'e when $2Q_1 > Q_2$ (a) and $2Q_1 < Q_2$ (b)} 
  \label{fig6}
\end{figure}

In Figure~\ref{fig5} one can see that the additional rows in the moir\'e pattern and the primary lattice are on the same side (bottom part of the images). This is at odds with what we observe in our experiments as shown in Figure~\ref{fig3}.a,d. There, the additional rows in the atomic and moir\'e lattices are located at the opposite sides of the dislocation. This peculiarity is explained in Figure~\ref{fig6}. Let us first assume that the additional row in the primary lattice is located at the bottom of the image as shown in Figure~\ref{fig5}.a,c. To first order, it corresponds to a term $\cos(Q_1x + \Delta\varphi(x,y))$. Then, the second order moir\'e pattern is given by the expression $\cos((2Q_1-Q_2)x + 2\Delta\varphi(x,y))$. In Figure~\ref{fig6} we show that depending on the sign of $(2Q_1-Q_2)$, the dislocation in the moiré pattern can appear either on the same side of the dislocation in the primary lattice ($2Q_1-Q_2 > 0$, Figure~\ref{fig6}.a), either in the opposite side ($2Q_1-Q_2 < 0$, Figure~\ref{fig6}.b). To demonstrate this, let us consider the phase term $(2Q_1-Q_2)x + 2\Delta\varphi(x,y)$ over a circuit surrounding the dislocation (see Figure~\ref{fig6}.a,b). In the case where $2Q_1 > Q_2$ (Figure~\ref{fig6}.a), along the blue arc, from A to B, the optical path difference $\Delta\theta$ is the sum of two positive contributions $\Delta\theta_1 = (2Q_1-Q_2)\Delta x + 2\pi$, while for the green arc, from B to A, the optical path difference is the sum of a negative term and a positive one $\Delta\theta_2 =-(2Q_1-Q_2)\Delta x + 2\pi$.
Hence, $|\Delta\theta_1|-|\Delta\theta_2| = 4\pi$, which means that the blue arc encompass two additional fringes as compared to the green arc. Now, in the case where $2Q_1 < Q_2$ (Figure~\ref{fig6}.b), the situation is just reversed. On the blue circuit the two phase shifts are opposite while on the green arc they add, this lead to the appearance of two additional rows in the top part of Figure~\ref{fig6}.b, while the in the primary lattice the dislocation is still located at the bottom (Figure~\ref{fig5}). The situation depicted in Figure~\ref{fig5}.b is in perfect agreement with what we observe in  Figure~\ref{fig5}.a, d and is consistent with the fact that $2Q_\textrm{CrCl3} < Q_\textrm{Au}$. The double dislocation in CrCl\textsubscript{3}/Au(111) is carrying a Berry phase of $-4\pi$, which correspond to a Chern winding number of 2, which is equivalent to a double vortex. 

\section{Conclusion \& Outlook}\label{sec3}

We have observed a higher order moir\'e pattern in a monolayer of CrCl\textsubscript{3} on a Au(111) surface that is driven by a large lattice mismatch: two periods of Au(111) matching almost one period of CrCl$_3$. The moiré is revealed in STM topography images but it originates mainly from an electronic structure modulation due to a strong coupling to the substrate which is not expected in such a van der Waals material. This interaction with the substrate was ascertained by DFT calculations that reveal a substantial shift of the chemical potential in CrCl$_3$ as compared to a free standing monolayer. The in situ grown samples exhibit single edge dislocations in the CrCl$_3$ monolayer that manifest as a double dislocations in the moiré pattern. This doubling effect is a direct consequence of the second order moiré. The dislocations in the moir\'e pattern are structural topological defects which can be described, in the continuous limit, as double vortices carrying Berry phases of  $4\pi$. Our model can be used with many different systems and may serve to identify interesting ones that might host topological defects with a well controlled Chern number. 

Due to the substantial coupling to a substrate with a strong Rashba spin-orbit coupling, this system could be an interesting platform for the stabilization of exotic non-collinear magnetic textures induced by the periodic potential of the moir\'e. The dislocations/vortices observed in the moir\'e pattern could be potentially interesting to pin topological spin textures that may host non conventional excitations such as non abelian anyons. It has been recently proposed in CrBr$_3$/NbSe$_2$, that the moiré was responsible of the appearance of topological superconductivity in NbSe$_2$ \cite{Kezilebieke2022}. We may therefore expect that the topological defects in the moiré pattern of CrBr$_3$ could stabilize Majorana bound states.

\section{Method}\label{sec4}
\subsection{Experimental Section}\label{subsec4}
The experiments were performed under ultra high vacuum. The Au(111) single crystal was cleaned by cycles of argon-ion sputtering and annealing to 450°C. Starting from an anhydrous powder CrCl\textsubscript{3} was carefully degazed before being evaporated using a Knudsen cell on the clean Au(111) substrate at room temperature. To obtain flat CrCl\textsubscript{3} islands, the sample was post annealed to about 150°C for few minutes. STM experiments were realized in a home built low temperature STM and on a LT omicron commercial STM. Experiments were performed at 4K.

\subsection{Density functional theory calculations}\label{subsec4}
The vacuum LDOS calculations in Figure~\ref{fig: spectroscopy}.b have been done using the FLAPW method implemented in the code FLEUR \cite{FLEUR}. Shown are the two VacDOS calculations of a unsupported CrCl\textsubscript{3} monolayer (UML) and the CrCl\textsubscript{3} monolayer on top of a Au(111) surface. The presented VacDOS is calculated approximately $3\,\AA$ above the surface. Since the unit cell of the moir\'e pattern with its substrate is too large for being feasible in first principles calculations, the atomic positions of both Cr and Cl atoms are placed in a perfect hexagonal structure where, a lattice constant in between Au bulk and CrCl\textsubscript{3} has been applied (tests have shown negligible dependence of the VacDOS vs. small variations of the lattice parameter, distances and atomic positions of the van der Waals material). The lattice parameters and distances used are presented in Tab.~\ref{tab: DFT}
\begin{table}[h]
 \caption{Lattice parameters in $\AA$ used for vacuum LDOS in Figure~\ref{fig: spectroscopy}.b. $a_\textrm{in-plane}$ denotes the in-plane lattice constant of the hexagonal lattice, $\Delta d_\textrm{Cl-Cr}$ is the vertical distance between the Cl layers and the Cr layer, $\Delta d_\textrm{CrCl\textsubscript{3}-Au(111)}$ denotes the vertical distance between CrCl\textsubscript{3} and the Au(111) surface.}
  \begin{tabular}[htbp]{@{}llll@{}}
    \hline
    System & $a_\textrm{in-plane}$ ($\AA$) & $\Delta d_\textrm{Cl-Cr}$ ($\AA$) & $\Delta d_\textrm{CrCl\textsubscript{3}-Au(111)}$ ($\AA$)\\
    \hline
    CrCl\textsubscript{3} UML  & 5.909  & 1.289 & $-$ \\
    CrCl\textsubscript{3}/Au(111)   & 5.909 & 1.336 & 4.233 \\
    \hline
  \end{tabular} \label{tab: DFT}
\end{table}

\backmatter

\bmhead{Supplementary information}

The online version contains supplementary material available at

\bmhead{Acknowledgements}
We Acknowledged Johann Coraux for his thoughtful reading of the paper.

\section*{Declarations}

\begin{itemize}
\item Funding:
This work was supported by the French Agence Nationale de la Recherche through the contract ANR GINET2-0 (ANR-20-CE42-0011). Computing time was provided by the Consortium d’\'Equipements de Calcul Intensif (FRS-FNRS Belgium GA 2.5020.11) and the LUMI CECI/Belgium for awarding this project access to the LUMI supercomputer, owned by the EuroHPC Joint Undertaking, hosted by CSC (Finland) and the LUMI consortium through LUMI CECI/Belgium, ULiege-NANOMAT-SKYRM-1. Sebastian Meyer is a Postdoctoral Researcher [CR] of the Fonds de la Recherche Scientifique – FNRS. Bertrand Dup{\'e} is a Research Associate [CQ]  of the Fonds de la Recherche Scientifique – FNRS.
\item Conflict of interest/Competing interests: Not applicable
\item Ethics approval and consent to participate: Not applicable
\item Consent for publication: Not applicable
\item Data availability: Not applicable 
\item Materials availability: Not applicable
\item Code availability: Not applicable
\item Author contribution:
M.H. and T.C proposed and conceived the research, E.G., S.G., M.H., T.C, L.L., C.B. and F.S. performed the experiments with the technical assistance of P.D. and F.D., E.G., S.G. and M.H analyzed the data, E.G., T.C and M.H. performed the num\'erical simulation and derived the analytical model. S.M. and B.D. performed the DFT calculations. M.H., T.C and S.M. wrote the original manuscript. All the authors discussed the results. \end{itemize}

\bibliography{biblio_moire.bbl}

\end{document}